\documentclass[letterpaper, 10 pt, conference]{ieeeconf}

\IEEEoverridecommandlockouts                              

\usepackage{graphicx}
\usepackage{amsmath}
\usepackage{amsfonts}
\usepackage{amssymb}
\usepackage{subcaption}
\usepackage{float}
\usepackage{algpseudocode}
\usepackage{multirow}
\usepackage{nicefrac}
\usepackage{xurl}
\usepackage{cite}
\usepackage[table,xcdraw]{xcolor}
\usepackage{nicematrix,booktabs}
\usepackage{flushend}

\usepackage{listings}
\usepackage[framed,autolinebreaks,useliterate]{mcode}

\usepackage{letltxmacro}
\LetLtxMacro\orgvdots\vdots
\LetLtxMacro\orgddots\ddots

\makeatletter
\DeclareRobustCommand\vdots{%
  \mathpalette\@vdots{}%
}
\newcommand*{\@vdots}[2]{%
  \sbox0{$#1\cdotp\cdotp\cdotp\m@th$}%
  \sbox2{$#1.\m@th$}%
  \vbox{%
    \dimen@=\wd0 %
    \advance\dimen@ -3\ht2 %
    \kern.5\dimen@
    \dimen@=\wd2 %
    \advance\dimen@ -\ht2 %
    \dimen2=\wd0 %
    \advance\dimen2 -\dimen@
    \vbox to \dimen2{%
      \offinterlineskip
      \copy2 \vfill\copy2 \vfill\copy2 %
    }%
  }%
}
\DeclareRobustCommand\ddots{%
  \mathinner{%
    \mathpalette\@ddots{}%
    \mkern\thinmuskip
  }%
}
\newcommand*{\@ddots}[2]{%
  \sbox0{$#1\cdotp\cdotp\cdotp\m@th$}%
  \sbox2{$#1.\m@th$}%
  \vbox{%
    \dimen@=\wd0 %
    \advance\dimen@ -3\ht2 %
    \kern.5\dimen@
    \dimen@=\wd2 %
    \advance\dimen@ -\ht2 %
    \dimen2=\wd0 %
    \advance\dimen2 -\dimen@
    \vbox to \dimen2{%
      \offinterlineskip
      \hbox{$#1\mathpunct{.}\m@th$}%
      \vfill
      \hbox{$#1\mathpunct{\kern\wd2}\mathpunct{.}\m@th$}%
      \vfill
      \hbox{$#1\mathpunct{\kern\wd2}\mathpunct{\kern\wd2}\mathpunct{.}\m@th$}%
    }%
  }%
}
\makeatother

\usepackage[finalnew]{trackchanges}
\addeditor{TB}
\addeditor{NJ}
\addeditor{JK}
\addeditor{HP}

\newtheorem{defn}{Definition}

\title{\LARGE \bf
zonoLAB: A MATLAB Toolbox for Set-based Control \\ Systems Analysis Using Hybrid Zonotopes
}

\author{Justin Koeln$^{1}$, Trevor J. Bird$^{2}$, Jacob Siefert$^{3}$, Justin Ruths$^{1}$, Herschel C. Pangborn$^{3}$, and Neera Jain$^{2}$
\thanks{*This material is based upon work supported by the National Science Foundation Graduate Research Fellowship under Grant No. DGE-1333468.}
\thanks{$^{1}$Justin Koeln and Justin Ruths are with the Department of Mechanical Engineering,
        University of Texas at Dallas, Richardson, TX, USA
        {\tt\small justin.koeln@utdallas.edu}, {\tt\small jruths@utdallas.edu}}%
\thanks{$^{2}$Trevor J. Bird and Neera Jain are with the School of Mechanical Engineering,
        Purdue University,  West Lafayette, IN, USA
        {\tt\small bird6@purdue.edu},
        {\tt\small neerajain@purdue.edu}}%
\thanks{$^{3}$Jacob Siefert and Herschel C. Pangborn are with the Department of Mechanical Engineering,
        The Pennsylvania State University, University Park, PA, USA
        {\tt\small jas7031@psu.edu}, {\tt\small hcpangborn@psu.edu}}%
}

\begin{document}

\maketitle
\thispagestyle{empty}
\pagestyle{empty}

\begin{abstract}

This paper introduces zonoLAB, a MATLAB-based toolbox for set-based control system analysis using the hybrid zonotope set representation. Hybrid zonotopes have proven to be an expressive set representation that can exactly represent the reachable sets of mixed-logical dynamical systems and tightly approximate the reachable sets of nonlinear dynamic systems. Moreover, hybrid zonotopes can exactly represent the continuous piecewise linear control laws associated with model predictive control and the input-output mappings of neural networks with piecewise linear activation functions. The hybrid zonotope set representation is also highly exploitable, where efficient methods developed for mixed-integer linear programming can be directly used for set operation and analysis. The zonoLAB toolbox is designed to make these capabilities accessible to the dynamic systems and controls community, with functionality spanning fundamental operations with hybrid zonotope, constrained zonotope, and zonotope set representations, powerful set analysis tools, and general-purpose algorithms for reachability analysis of open- and closed-loop systems.

\end{abstract}


\section{Introduction}\label{sec-intro}

Set-based methods are increasingly used for the analysis and control of dynamic systems. These methods are supported by a well-established and rigorous theoretical foundation \cite{Blanchini2015book,Borrelli2011Book,Fukuda2016} and are becoming more practical in application due to expressive and efficient set representations \cite{Scott2016,Bird2023Automatica,Kochdumper2021a,Kousik2023,Silvestre2022}, improved algorithms \cite{Sadraddini2019b,Raghuraman2020,Yang2018}, and the availability of open-source software packages \cite{Althoff2020,Kvasnica2015,Bak2017,Frehse2011b,Meyer2019,Bogomolov2019,Forets2021,Vinod2019}. Within the field of set-based methods, a subset of techniques are based on set propagation and reachability analysis, as reviewed in \cite{Althoff2021}. Whether used for safety verification, state estimation, dynamic system analysis, or control design, set propagation techniques generally focus on iterative procedures to determine the set of states reachable by a dynamic system, given a bounded set of initial conditions and bounded sets of controllable inputs, exogenous disturbances, and/or model uncertainty. Open-source software toolboxes are available for a variety of set-based methods \cite{Althoff2020,Kvasnica2015,Bak2017,Frehse2011b,Meyer2019,Bogomolov2019,Forets2021,Vinod2019}. 
While a complete review of available toolboxes falls outside the scope of this brief paper, each toolbox is tailored to balance performance and computational cost when applying select analysis techniques to particular classes of systems through the choice of appropriate set representations and algorithms that exploit the structure of these representations.  

The objective of this paper is to introduce zonoLAB\footnotemark, a MATLAB toolbox that strategically exploits the structures of zonotopic set representations, including zonotopes, constrained zonotopes, and hybrid zonotopes, for a wide range of set operations and applications. A unique feature of zonoLAB is the use of hybrid zonotopes, a set representation using $ n_b $ binary factors to define the union of up to $ 2^{n_b} $ constrained zonotopes (i.e., convex polytopes).
In addition to basic set operations, such as linear mappings, Minkowski sums, and intersections, zonoLAB provides user-friendly implementations of recent results exploiting the hybrid zonotope set representation, such as reachability analysis of Mixed Logical Dynamical (MLD) and Piecewise Affine (PWA) systems \cite{Bird2023Automatica}, set-based representations of Model Predictive Control (MPC) feedback policies and closed-loop reachability \cite{Bird2022ACC}, exact set-based representations of neural networks with ReLU activations functions \cite{Ortiz2023}, and set-based approximations of nonlinear system dynamics for analysis and control design \cite{Siefert2023TAC}.   

\footnotetext{Available for download at \url{https://github.com/ESCL-at-UTD/zonoLAB} under the GNU General Public License.}


 
The remainder of this paper is organized as follows. Section \ref{sec-prelims} provides background information on zonotopic set representations and operations. The structure and main functionality of zonoLAB are presented in Section \ref{sec-fundamentals}. Section \ref{sec-examples} demonstrates how zonoLAB can be used to compute reachable sets for MLD and PWA systems, MPC-controlled systems, neural networks, and systems with nonlinear dynamics. Section \ref{sec-conclusions} concludes the paper.

\noindent\textbf{Notation:} 
Sets are denoted by uppercase calligraphic letters, e.g., $\mathcal{Z}\subset\mathbb{R}^{n}$. Commas in subscripts are used to distinguish between properties that are defined for multiple sets; e.g., $n_{g,z}$ describes the complexity of the set $\mathcal{Z}$. The $n$-dimensional unit hypercube is denoted by $\mathcal{B}_{\infty}^n=\left\{x\in\mathbb{R}^{n}~|~\|x\|_{\infty}\leq1\right\}$ while the constrained unit hypercube is $ \mathcal{B}_\infty^{n_g}(A,b) =\left\{x\in\mathbb{R}^{n}~|~\|x\|_{\infty} \leq 1, A x = b \right\} $. The set of all $n$-dimensional binary vectors is denoted by $\{-1,1\}^{n}$.
The cardinality of the discrete set $\mathcal{T}$ is denoted by $|\mathcal{T}|$.
The concatenation of two column vectors to a single column vector is denoted by $(\xi_1~\xi_2)=[\xi_1^\top~\xi_2^\top]^\top$. The bold numbers $\mathbf{1}$ and $\mathbf{0}$ denote matrices of all $1$ and $0$ elements, respectively, and $\mathbf{I}$ denotes the identity matrix with dimensions indicated by subscripts when not easily deduced from context. 
\section{Set Representations and Basic Operations}\label{sec-prelims}

\subsection{Zonotopic Set Representations}

The zonoLAB toolbox is primarily built on three main zonotopic representations: the zonotope, the constrained zonotope, and the hybrid zonotope.

\begin{defn} (Zonotope) \cite{McMullen1971}
A set $ \mathcal{Z} \subset \mathbb{R}^n $ is a \emph{zonotope} in Generator-representation (G-rep) if there exists generator matrix $ G \in \mathbb{R}^{n \times n_g } $ and center $ c \in \mathbb{R}^n $ such that 
\begin{equation}
    \mathcal{Z} = \{ G \xi + c \; | \; \|\xi\|_\infty \leq 1 \},
\end{equation}
and is denoted by $ \mathcal{Z} = \langle G, c \rangle $.
\end{defn}

\begin{defn} (Constrained Zonotope) \cite{Scott2016}
A set $ \mathcal{Z}_c \subset \mathbb{R}^n $ is a \emph{constrained zonotope} in Constrained Generator-representation (CG-rep) if there exists a generator matrix $ G \in \mathbb{R}^{n \times n_g } $, center $ c \in \mathbb{R}^n $, constraint matrix $ A \in \mathbb{R}^{n_c \times n_g}$, and constraint vector $ b \in \mathbb{R}^{n_c} $ such that 
\begin{equation}
    \mathcal{Z}_c = \{ G \xi + c \; | \; \|\xi\|_\infty \leq 1, A \xi = b \},
\end{equation}
and is denoted by $ \mathcal{Z}_c = \langle G, c, A, b \rangle $.
\end{defn}

\begin{defn} (Hybrid Zonotope) \cite{Bird2023Automatica}
A set $ \mathcal{Z}_h \subset \mathbb{R}^n $ is a \emph{hybrid zonotope} in Hybrid Constrained Generator-representation (HCG-rep) if there exists a continuous generator matrix $ G^c \in \mathbb{R}^{n \times n_g } $, binary generator matrix $ G^b \in \mathbb{R}^{n \times n_b } $, center $ c \in \mathbb{R}^n $, continuous constraint matrix $ A^c \in \mathbb{R}^{n_c \times n_g}$, binary constraint matrix $ A^b \in \mathbb{R}^{n_c \times n_b}$, and constraint vector $ b \in \mathbb{R}^{n_c} $ such that 
\begin{equation}
        \mathcal{Z}_h = \left\{ \left[G^c \: G^b\right]\left[\begin{smallmatrix}\xi^c \\ \xi^b \end{smallmatrix}\right]  + c\: \middle| \begin{matrix} \left[\begin{smallmatrix}\xi^c \\ \xi^b \end{smallmatrix}\right]\in \mathcal{B}_\infty^{n_{g}} \times \{-1,1\}^{n_{b}}, \\ \left[A^c \: A^b\right]\left[\begin{smallmatrix}\xi^c \\ \xi^b \end{smallmatrix}\right] = b \end{matrix} \right\},
\end{equation}
and is denoted by $ \mathcal{Z}_h = \langle G^c, G^b, c, A^c, A^b, b \rangle $.
\end{defn}

For zonotopes and constrained zonotopes, the columns of $ G $ are called the \emph{generators} and the elements of $ \xi \in \mathbb{R}^{n_g} $ are called \emph{factors}. For hybrid zonotopes, the columns of $ G^c $ ($ G^b $) are called the \emph{continuous generators} (\emph{binary generators}) and the elements of $ \xi^c \in \mathbb{R}^{n_g} $ ($ \xi^b \in \mathbb{R}^{n_b} $) are called \emph{continuous factors} (\emph{binary factors}).

\begin{figure}[t]
     \centering
     \includegraphics[width=0.9\linewidth]{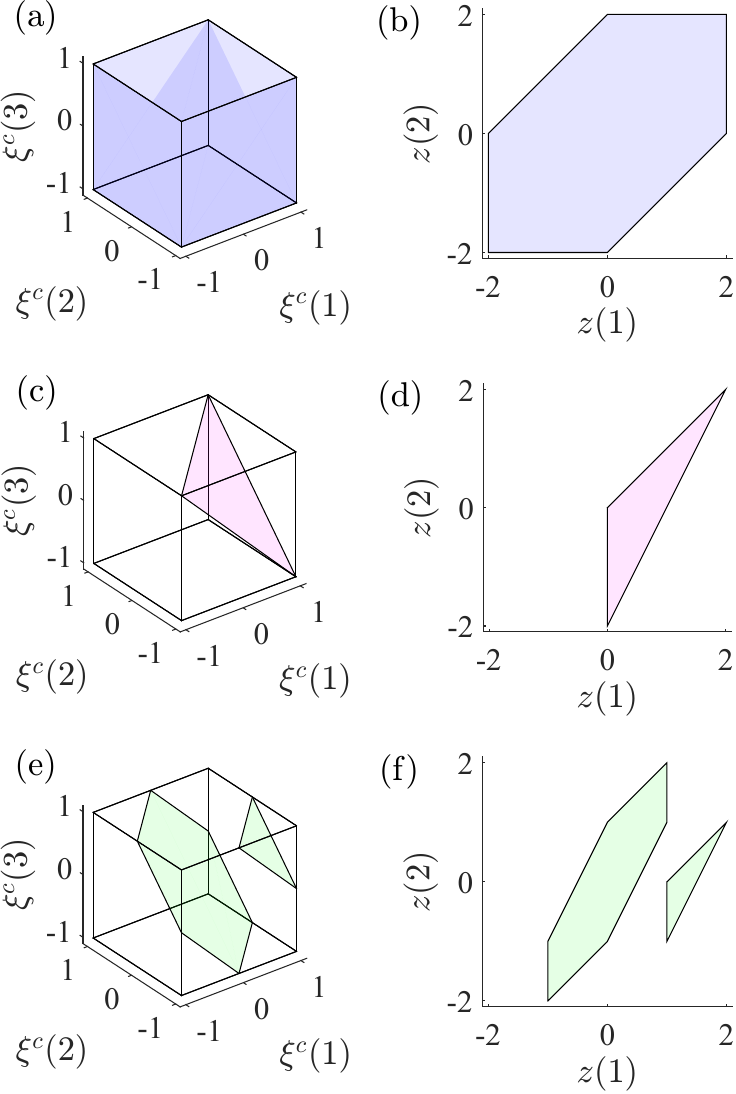}
     \caption{Examples of how the constrained factor spaces (a), (c), and (e) are projected to create a zonotope (b), constrained zonotope (d), and hybrid zonotope (f), respectively, generated by the \mcode{exampleSetDefinition.m} file provided with zonoLAB.}
    \label{fig:setDefinition}
\end{figure}

As shown in Fig. \ref{fig:setDefinition}, a zonotope is an affine image of the unit hypercube $ \mathcal{B}_\infty^{n_g} $, while a constrained zonotope is an affine image of the constrained unit hypercube $ \mathcal{B}_\infty^{n_g}(A,b) $. By introducing $n_b$ binary factors, a hybrid zonotope is equivalent to the union of $|\mathcal{T}|\leq2^{n_b}$ constrained zonotopes as
\begin{subequations}\label{thm-eqn-decomposedZc}
\begin{align}
    &\mathcal{Z}_{h}=\bigcup_{\xi^b_i\in\mathcal{T}}\mathcal{Z}_{c,i}\:, \label{thm-eqn-decomposedZc-complex}\\
    &\mathcal{Z}_{c,i}=\left\langle G^c,c+G^b\xi^b_i,A^c,b-A^b\xi^b_i\right\rangle\:, \label{thm-eqn-decomposedZc-CG}
\end{align}
\end{subequations}
where the $i^{th}$ constrained zonotope \eqref{thm-eqn-decomposedZc-CG} has its center and constraint vector shifted by the $i^{th}$ value of the discrete set $\mathcal{T}=\{\xi_i^b\in\{-1,1\}^{n_{b}}~|~\mathcal{Z}_{c,i}\not=\emptyset\}$ mapped by $G^b$ and $A^b$, respectively. Therefore, while a constrained zonotope is the affine image of the intersection of a hypercube and hyperplanes, a hybrid zonotope is the affine image of multiple intersections of a hypercube and shifted  hyperplanes, as shown in Fig. \ref{fig:setDefinition}.

Note that zonotopes are a subset of convex polytopes that are centrally symmetric, constrained zonotopes can represent any convex polytope \cite{Scott2016}, and hybrid zonotopes can represent the union of any collection of convex polytopes \cite{Bird2023Automatica}. Moreover, zonotopes are a subset of constrained zonotopes, which are a subset of hybrid zonotopes, and a hybrid zonotope degenerates to a constrained zonotope when $ n_b = 0 $, which degenerates to a zonotope when $ n_c = 0 $.

\begin{table*}
     \centering
     \caption{Set operations provided with the zonoLAB toolbox.\protect\footnotemark \space Computational complexity assumes sets in $n$-dimensions and $n_g$ generators (where $n_g \leftarrow n_g + n_b$ for hybrid zonotopes) and matrix $R \in \mathbb{R}^{m \times n}$.}
    \label{fig-methodTable}
     \begin{tabular}{llll}
\hline
\multicolumn{1}{|c|}{\textbf{Operation}}                                           & \multicolumn{1}{c|}{\textbf{Mathematically}}                                                        & \multicolumn{1}{c|}{\textbf{Method}}                                      & \multicolumn{1}{c|}{\textbf{Scalability}}                               \\ \hline
\multicolumn{1}{|l|}{Generalized intersection}                                     & \multicolumn{1}{l|}{$\mathcal{Z} \cap_R \mathcal{Y}$}                                               & \multicolumn{1}{l|}{\mcode{and} or \mcode{\&}}                                & \multicolumn{1}{l|}{$\mathcal{O}(m n n_g)$}                             \\ \hline
\rowcolor[HTML]{EFEFEF} 
\multicolumn{1}{|l|}{\cellcolor[HTML]{EFEFEF}Cartesian product}                    & \multicolumn{1}{l|}{\cellcolor[HTML]{EFEFEF}$\mathcal{Z} \times \mathcal{Y}$}                       & \multicolumn{1}{l|}{\cellcolor[HTML]{EFEFEF}\mcode{cartProd}}               & \multicolumn{1}{l|}{\cellcolor[HTML]{EFEFEF}$\mathcal{O}(1)$}           \\ \hline
\multicolumn{1}{|l|}{Convex hull$^a$}                                              & \multicolumn{1}{l|}{$\text{CH}(\mathcal{Z} \cup \mathcal{Y}) $}                                     & \multicolumn{1}{l|}{\mcode{convexHull}}                                     & \multicolumn{1}{l|}{$\mathcal{O}(n_g)$}                                 \\ \hline
\rowcolor[HTML]{EFEFEF} 
\multicolumn{1}{|l|}{\cellcolor[HTML]{EFEFEF}Intersection with halfspace(s)}       & \multicolumn{1}{l|}{\cellcolor[HTML]{EFEFEF}$\mathcal{Z} \cap_R \lbrace x \mid H x \leq f \rbrace$} & \multicolumn{1}{l|}{\cellcolor[HTML]{EFEFEF}\mcode{halfspaceIntersection}} & \multicolumn{1}{l|}{\cellcolor[HTML]{EFEFEF}$\mathcal{O}(n_H m n n_g)$} \\ \hline
\multicolumn{1}{|l|}{Linear mapping}                                               & \multicolumn{1}{l|}{$R \mathcal{Z} $}                                                               & \multicolumn{1}{l|}{\mcode{mtimes} or \mcode{*}}                             & \multicolumn{1}{l|}{$\mathcal{O}(m n n_g)$}                             \\ \hline
\rowcolor[HTML]{EFEFEF} 
\multicolumn{1}{|l|}{\cellcolor[HTML]{EFEFEF}Minkowski sum}                        & \multicolumn{1}{l|}{\cellcolor[HTML]{EFEFEF}$\mathcal{Z} \oplus \mathcal{Y}$}                       & \multicolumn{1}{l|}{\cellcolor[HTML]{EFEFEF}\mcode{plus} or \mcode{+}}        & \multicolumn{1}{l|}{\cellcolor[HTML]{EFEFEF}$\mathcal{O}(n)$}           \\ \hline
\multicolumn{1}{|l|}{Pontryagin difference$^b$}                                      & \multicolumn{1}{l|}{$\mathcal{Z} \ominus \mathcal{Y}$}                                              & \multicolumn{1}{l|}{\mcode{pontryDiff}}                                     & \multicolumn{1}{l|}{$\mathcal{O}(2^{n_g} n^2 n_g)$}                      \\ \hline
\rowcolor[HTML]{EFEFEF} 
\multicolumn{1}{|l|}{\cellcolor[HTML]{EFEFEF}Projection onto subset of dimensions} & \multicolumn{1}{l|}{\cellcolor[HTML]{EFEFEF}$\Pi_d \mathcal{Z}$}                                    & \multicolumn{1}{l|}{\cellcolor[HTML]{EFEFEF}\mcode{projection}}             & \multicolumn{1}{l|}{\cellcolor[HTML]{EFEFEF}$\mathcal{O}(1)$}           \\ \hline
\multicolumn{1}{|l|}{One-step reach set for MLD system}                            & \multicolumn{1}{l|}{$\mathcal{R}_k \rightarrow \mathcal{R}_{k+1} $}                               & \multicolumn{1}{l|}{\mcode{stepMLD}}                                        & \multicolumn{1}{l|}{$\mathcal{O}(n^3_g)$}                               \\ \hline
\rowcolor[HTML]{EFEFEF} 
\multicolumn{1}{|l|}{\cellcolor[HTML]{EFEFEF}Union}                                & \multicolumn{1}{l|}{\cellcolor[HTML]{EFEFEF}$\mathcal{Z} \cup \mathcal{Y} $}                        & \multicolumn{1}{l|}{\cellcolor[HTML]{EFEFEF}\mcode{union}}                  & \multicolumn{1}{l|}{\cellcolor[HTML]{EFEFEF}$\mathcal{O}(n_g)$}         \\ \hline
\multicolumn{4}{l}{$^a$ For \mcode{zono} and \mcode{conZono} objects only. $\quad$ $^b$ Set $\mathcal{Y}$ must be a \mcode{zono} object.}                                                  
\end{tabular}
\vspace{-5pt}
\end{table*}

\subsection{Zonotopic Set Operations}


Historically, the computational benefits of using a zonotope set representation come from the relative ease, numerical accuracy, and scalability of computing linear mappings and Minkowski sums, where given matrix $ R \in \mathbb{R}^{m \times n} $ and zonotopes $ \mathcal{Z},\mathcal{W} \subset \mathbb{R}^n $,
\begin{subequations}
\begin{align}
    R\mathcal{Z} =& \langle R G_z, R c_z \rangle \:, \\
    \mathcal{Z} \oplus \mathcal{W} =& \langle [ G_z G_w ], c_z + c_w \rangle \:.
\end{align}
\end{subequations}

Such practical benefits were extended to the general class of convex polytopes with the invention of \emph{constrained zonotopes} \cite{Scott2016}. While linear mappings and Minkowski sums are useful for reachability analysis, additional operations are also made computationally efficient through the use of zonotopic set operations. For example, while it has been shown that the zonotope containment problem is co-NP-complete \cite{Kulmburg2021}, the necessary, but not sufficient, condition for $ \mathcal{Z} \subseteq \mathcal{W} $ if $ \mathcal{Z} $ and $ \mathcal{W} $ are zonotopes is formulated as a feasibility check for a system of linear equations. This enabled set containment conditions to be directly embedded in robust MPC formulations \cite{Raghuraman2021ACC}. The zonoLAB toolbox aims to provide a unified framework that implements recent developments in zonotope, constrained zonotope, and hybrid zonotope set operations \cite{Raghuraman2020,Bird2023Automatica,Bird2022ACC,Bird2021,Siefert2023ACC,Siefert2022,Siefert2023TAC,Ortiz2023} to make the practical benefits of these set representations more accessible to the dynamic systems and controls community. 


\footnotetext{The similar table at \url{https://github.com/ESCL-at-UTD/zonoLab/wiki/Functions} will be updated if future releases include additional set operations or different computational complexities.}
\section{The zonoLAB Toolbox }\label{sec-fundamentals}


\subsection{Toolbox Structure}

Using Object-Oriented Programming (OOP) in MATLAB, zonoLAB consists of three main classes---\mcode{zono}, \mcode{conZono}, and \mcode{hybZono}---used to represent zonotopes, constrained zonotopes, and hybrid zonotopes, respectively. An abstract class, \mcode{abstractZono}, is used as a superclass for these three classes. As an abstract class, \mcode{abstractZono} cannot be instantiated but is used to define common functionality for the \mcode{zono}, \mcode{conZono}, and \mcode{hybZono} classes.

Methods provided by the zonoLAB toolbox are organized into five categories: properties, arithmetic, visualization, complexity, and auxiliary. For brevity, Table \ref{fig-methodTable} provides a list of only the arithmetic methods currently available using the three zonotopic set representations, along with the computational complexity of each operation. These methods are defined for the \mcode{abstractZono} superclass to be used with any of the three main classes. 

\subsection{Set Definitions and Conversions} \label{sec:setDefn}

The hybrid zonotope $ \mathcal{Z}_h = \langle G^c, G^b, c, A^c, A^b, b \rangle $ is defined using zonoLAB as
\vspace{-\baselineskip}
\begin{lstlisting} 
Zh = hybZono(Gc,Gb,c,Ac,Ab,b)
\end{lstlisting}
assuming the variables \mcode{Gc,Gb,c,Ac,Ab,b} are predefined. Zonotopes and constrained zontopes are defined similarly. Sets can also be recast to a more expressive representation. For example, a constrained zonotope, \mcode{Zc}, can be recast as a hybrid zonotope using
\vspace{-\baselineskip}
\begin{lstlisting}
Zh = hybZono(Zc)
\end{lstlisting}
which automatically sets \mcode{Zh.Gb} and \mcode{Zh.Ab} to matrices of all zeros of the appropriate dimension. 

A convex polytopic set defined in H-rep as $ \mathcal{H} = \{ x \in \mathbb{R}^n | H x \leq f \} $, where $ H \in \mathbb{R}^{n_h \times n } $, $ f \in \mathbb{R}^{n_h} $, and $ n_h $ is the number of halfspaces, can be represented as a constrained zonotope using
\vspace{-\baselineskip}
\begin{lstlisting}
H_CG = conZono([H f])
\end{lstlisting}

The union of $ N $ sets in H-rep can be represented as a hybrid zonotope using
\vspace{-\baselineskip}
\begin{lstlisting}
H_HCG = hybZono(H_collection)
\end{lstlisting}
where \mcode{H_collection} is a $ N \times 1 $ cell array where the $ i^{th} $ cell is $ [H_i \; f_i] $ corresponding to the $ i^{th} $ set $ \mathcal{H}_i = \{ x \in \mathbb{R}^n | H_i x \leq f_i \} $. Note that the $ N $ sets do not need to have the same number of halfspaces.

For a collection of $ n_v $ vertices $\{v_1, \dots, v_{n_v} \} $, where $ v_i \in \mathbb{R}^n $, a set $ \mathcal{V} $ defined in V-rep is conventionally considered as the convex hull of the vertices, $ \mathcal{V} = \text{conv}(\{v_1, \dots, v_{n_v} \}) $. However, zonoLAB provides the flexibility to use this collection of vertices to define a set as the union of the vertices, as the union of line segments between vertices, as the conventional convex hull of these vertices, or some combination of these three \cite{Siefert2023TAC}. Thus, in addition to providing the vertex matrix $ V = [ v_1, \dots, v_{n_v} ] \in \mathbb{R}^{n \times n_v} $, an incidence matrix $ M \in \mathbb{R}^{n_v \times N}$ must also be provided, where the rows correspond to the $n_v$ vertices and the columns correspond to the desired $ N $ convex sets. Elements $ M_{i,j} = 1 $ if the $ i^{th} $ vertex is a member of the $ j^{th} $ set. All other entries are zero. For example, if $ M = \mathbf{I}_{n_v} $, the result is a hybrid zonotope representing the union of the $ n_v $ vertices. If $ M $ is defined such that $ M_{1,1} = M_{1,N} = 1 $ and $ M_{i,i-1} = M_{i,i} = 1, \forall i \in \{2,\dots,n_v\} $, then the result is a hybrid zonotope representing the line segments between each consecutive vertex in the vertex matrix, including a line segment between the first and last vertices. Finally, if $ M = \mathbf{1}_{n_v \times 1 } $, then the result is a constrained zonotope representing the convex hull of the vertices. In general, with matrices \mcode{V} and \mcode{M} defined, the desired set is created using
\vspace{-\baselineskip}
\begin{lstlisting}
V_HCG = hybZono({V,M})
\end{lstlisting}
where $ \mcode{\{V,M\}} $ is a $ 1 \times 2 $ cell array.

\subsection{Plotting} \label{sec:plotting}

In zonoLAB, plotting a zonotopic set is achieved by identifying all vertices of the set, and the corresponding collection of faces composed of subsets of these vertices, to use MATLAB's \mcode{patch} function for visualizing 2D and 3D polygons. As a method of the \mcode{abstractZono} class, \mcode{[v,f] = plot(Z,optPlot)} is used to determine the $ n_v $ vertices and $n_f$ faces of the $n$-dimensional set \mcode{Z} stored in the $ n_v \times n $ matrix \mcode{v} and the $ n_f \times n_{max} $ matrix \mcode{f}, where $ n_{max} $ is the largest number of vertices on a single face. The second argument \mcode{optPlot} is an object of the \mcode{plotOptions} class used to tailor display properties for the set. Since plotting \mcode{conZono} and \mcode{hybZono} objects requires solving linear programs, a \mcode{solverOptions} class is also included in zonoLAB for users to specify their preferred solver for each type of optimization problem. The \mcode{plot} method calls set-specific methods for the \mcode{zono}, \mcode{conZono}, and \mcode{hybZono} classes. For brevity, only the main ideas of the 3D plotting algorithms provided in zonoLAB are presented here, assuming that the set is full-dimensional. 

For \mcode{zono} objects, zonoLAB exploits the fact that, in 3D, the faces of a zonotope are parallelograms formed by two of the generators from the zonotope centered at a linear combination of the remaining $n_g-2$ generators \cite{Guibas2003}. As a result, the number of vertices of a 3D zonotope is bounded such that $ n_v \leq 2 \sum_{i=0}^{2} \binom {n_g-1} i $ and the number of faces is bounded such that $n_f \leq 2 \binom {n_g} 2 $ \cite{Fukuda2016}. However, note that the number of rows of \mcode{v} will likely exceed this bound on $ n_v $ since no effort is made to identify and remove repeated vertices. Since plotting \mcode{zono} objects does not require solving linear programs, plotting sets is relatively fast even when $n_g$ is large (e.g., a random zonotope $ \mathcal{Z}\in\mathbb{R}^3$ with $ n_g = 200 $ produces $ n_v = 159,200 $ vertices and $ n_f = 39,800 $ faces and is plotted in approximately 1 second on a standard desktop computer). 

Plotting \mcode{conZono} and \mcode{hybZono} objects is considerably more time intensive. The zonoLAB toolbox uses a vertex-centric approach, where linear programming is used to identify each vertex and the number of optimization problems to be solved scales linearly with the number of vertices in the set. Currently, zonoLAB plots \mcode{hybZono} objects by plotting each constrained zonotope from \eqref{thm-eqn-decomposedZc-CG} and using the \mcode{getLeaves} method to determine the discrete set $\mathcal{T}$ containing the combinations of binary factors that result in non-empty constrained zonotopes. Therefore, plotting a \mcode{hybZono} object is the same as plotting $|\mathcal{T}|$ \mcode{conZono} objects. 

The zonoLAB toolbox plots \mcode{conZono} objects in 3D using a modified version of the expanding polytope algorithm from \cite{VandenBergen2001}. With the goal of finding all vertices of a convex polytope, linear programming is used to determine the furthest point in a particular direction while remaining within the set. After identifying an initial 3-simplex with 4 vertices, normal vectors for each face are used as these search directions. If a vertex is identified that is further in this direction than the existing face, the vertex is added to the list of vertices, the existing face is removed from the list of faces, and new faces are defined that include this new vertex. This process is repeated until all face normals have been used as search directions and no new vertices are found. While slower than \mcode{zono} plotting, this approach remains practical for medium-complexity sets (e.g., a random constrained zonotope $ \mathcal{Z}_c\in\mathbb{R}^3$ with $ n_g = 50 $ and $ n_c = 26 $, produces $ n_v = 7,668 $ vertices and $ n_f = 7,762 $ faces and is plotted in approximately 60 seconds on a standard desktop computer, where GUROBI \cite{Gurobi} is called to solve a linear program 15,432 times).

\subsection{Set-based Mappings} \label{sec:setBasedMap}

The zonoLAB toolbox also leverages the notion of a \emph{set-based mapping}, where dimensions of a set are categorized as either \emph{input} or \emph{output} dimensions and the set provides the mapping between inputs and outputs over a bounded input domain. This notation is similar to the \emph{graph of a function}, where if $ f: \mathcal{X} \rightarrow \mathcal{Y} $ then the graph is $ G(f) = \{(x,f(x)) \, | \, x \in \mathcal{X} \} \subseteq \mathcal{X} \times \mathcal{Y} $. The idea of set-based mapping is demonstrated using the following linear system example but is generalizable (and more valuable) when representing piecewise linear and nonlinear mappings.

Consider a linear, discrete-time, time-invariant system $ x_{k+1} = A x_k + B u_k $, where $ x_k \in \mathcal{R}_k \subset \mathbb{R}^n $, $ u_k \in \mathcal{U} \subset \mathbb{R}^m $, and $ A, B, \mathcal{R}_k $, and $ \mathcal{U} $ are known. The one-step reachable set $\mathcal{R}_{k+1}$ is computed as
\begin{equation} \label{eq:oneStepReachExample}
    \mathcal{R}_{k+1} = A \mathcal{R}_k \oplus B \mathcal{U}\:.
\end{equation}

Using zonoLAB, with \mcode{A}, \mcode{B}, \mcode{Rk}, and \mcode{U} predefined, the reachable set \mcode{Rkplus} can be computed using methods \mcode{mtimes} and \mcode{plus} of the \mcode{abstractZono} class as
\vspace{-\baselineskip}
\begin{lstlisting}
Rkplus = A*Rk + B*U
\end{lstlisting}

The one-step reachable set $\mathcal{R}_{k+1}$ is also known as the \emph{successor set}, $ \text{Suc}(\mathcal{R}_k,\mathcal{U})$, which can be defined for the more general nonlinear dynamics $ x_{k+1} = f(x_k, u_k) $ as
\begin{equation}
    \text{Suc}(\mathcal{R}_k,\mathcal{U}) = \{ f(x_k,u_k) \; | \; x_k \in \mathcal{R}_k, u_k \in \mathcal{U} \}\:.
\end{equation}

The notion of an \emph{open-loop state-update set} was presented in \cite{Siefert2023TAC}, where a set is used to bound all possible state transitions given by $ f(\cdot,\cdot) $ over a predefined bounded domain of states and inputs. These state-update sets are an example of a set-based mapping. When a state-update set capturing open-loop dynamics is coupled to a similar set-based map of a given control law, the resulting \emph{closed-loop state-update set} bounds the evolution of the closed-loop system and can be used for closed-loop reachability analysis and verification. 

Mathematically, the open-loop state-update set $  \Psi \subset \mathbb{R}^{2n+m} $ is defined as
\begin{equation} \label{eq:openLoopStateUpdate}
    \Psi = \Bigg \{ \begin{bmatrix} x_k \\ u_k \\ x_{k+1} \end{bmatrix}\ \bigg |\ \begin{array}{c} x_{k+1} \in \text{Suc} (\{x_{k}\},\{u_k\}),\\ (x_k,u_k) \in \text{D}(\Psi) \end{array} \Bigg \}\:,
\end{equation}
where $ \text{D}(\Psi) \subset \mathbb{R}^{n+m} $ is the \emph{domain set} of $ \Psi $, which is chosen as a region of interest for analysis, as in \cite{Siefert2022}. 

Using the open-loop state-update set $ \Psi $, with domain set defined such that $ \mathcal{R}_k \times \mathcal{U} \subseteq \text{D}(\Psi) $, the successor set can be computed using projection and generalized intersection operations as
\begin{equation} \label{eq:successorSet}
    \text{Suc}(\mathcal{R}_k,\mathcal{U}) = \begin{bmatrix}
        \mathbf{0}_{n \times n+m} & \textbf{I}_n
        \end{bmatrix}\big(\Psi \cap_{[\textbf{I}_{n+m}~\textbf{0}_n]} (\mathcal{R}_k \times \mathcal{U})\big)\:.
\end{equation}

While \eqref{eq:oneStepReachExample} provides a method for computing the one-step reachable set, a state-update set  $ \Psi $ can also be used to compute $ \mathcal{R}_{k+1} $. Assuming the domain set $ \text{D}(\Psi) \in \mathbb{R}^{n+m} $ is known such that $ \mathcal{R}_k \times \mathcal{U} \subseteq \text{D}(\Psi) $, then 
\begin{equation} \label{eq:stateUpdateSet}
    \Psi = \begin{bmatrix}
        \mathbf{I}_n & \mathbf{0}_{n \times m} \\ \mathbf{0}_{m \times n} & \mathbf{I}_m \\ A & B 
    \end{bmatrix} 
    \text{D}(\Psi)\:.
\end{equation}
This state-update set and the successor set operation from \eqref{eq:successorSet} can be used to compute $ \mathcal{R}_{k+1} $ using zonoLAB as
\vspace{-\baselineskip}
\begin{lstlisting}
D = zono(G_d, c_d)
Psi = [eye(n+m); A B]*D
Rkplus = [zeros(n) eye(n)]* and(Psi,cartProd(Rk,U),[eye(n+m) zeros(n)])
\end{lstlisting}
When comparing the use of \eqref{eq:successorSet} and \eqref{eq:stateUpdateSet} with the use of \eqref{eq:oneStepReachExample}, it may appear that there are considerable disadvantages with the state-update set approach, namely $ \mathcal{R}_{k+1} $ is only computed accurately if  $ \mathcal{R}_k \times \mathcal{U} \subseteq \text{D}(\Psi) $ and the resulting set is a \mcode{conZono} (due to the use of generalized intersection in \eqref{eq:successorSet}) instead of the \mcode{zono} object obtained when using \eqref{eq:oneStepReachExample} (assuming $ \mathcal{R}_k $ and $ \mathcal{U} $ are \mcode{zono} objects). However, the use of state-update sets (and set-based mappings in general) allows reachability analysis based on \eqref{eq:successorSet} to be easily extended to hybrid systems (including MLD and PWA systems) and nonlinear systems. Additionally, as discussed in \cite{Siefert2022}, state-update sets can also be used to compute precursor sets for backward reachability analysis. The following sections demonstrate how set-based mappings (including state-update sets) provide a unifying framework for several different applications of reachability analysis. 


\section{Reachability Analysis Examples} \label{sec-examples}

The purpose of this section is to demonstrate how zonoLAB can be used to conduct reachability analysis of several different types of systems using the unique capabilities of the hybrid zonotope representation. 

\subsection{MLD and PWA System Analysis}\label{sec-MLD}

Based on the work from \cite{Bird2023Automatica}, zonoLAB can compute the reachable sets of MLD systems, a modeling framework combining continuous and binary variables with logical relations in mixed-integer inequalities to express complex dynamic systems \cite{Bemporad1999}. A linear discrete-time MLD system is expressed as 
\begin{subequations}\label{eqn-MLDdef}
\begin{align}
x_{k+1}=A&x_k+B_{u}u_k+B_{w}w_k+B_{aff}\:,\label{eqn-MLDdef-state}\\
\text{s.t.}\:\:\:E_{x}&x_k+E_{u}u_k+E_{w}w_k\leq E_{aff}\:,\label{eqn-MLDdef-inequality}
\end{align}
\end{subequations}
where $x\in \mathbb{R}^{n_{xc}}\times\{0,1\}^{n_{xl}}$ are the system states, $u\in \mathbb{R}^{n_{uc}}\times\{0,1\}^{n_{ul}}$ are the control inputs, and $w \in \mathbb{R}^{n_{rc}}\times\{0 , 1\}^{n_{rl}}$ are auxiliary variables. The number of inequality constraints is denoted by $n_e$ such that $E_{aff}\in \mathbb{R}^{n_e}$. Assuming knowledge of \mcode{A}, \mcode{B_u}, \mcode{B_w}, \mcode{B_aff}, \mcode{E_x}, \mcode{E_u}, \mcode{B_w}, and \mcode{E_aff} along with the sets \mcode{Rk} and \mcode{U} (such that $ x_k \in \mathcal{R}_k$ and $ u_k \in \mathcal{U}$), the compact set \mcode{W} (such that $w_k \in \mathcal{W} \subset {\mathbb{R}^{n_{rc}}}\times\{0,1\}^{n_{rl}}$ must also be provided.\footnotemark \space Using zonoLAB, the successor set for the MLD system is computed as 
\vspace{-\baselineskip}
\begin{lstlisting}
Rkplus = stepMLD(Rk,U,W,A,B_u,B_w,B_aff, E_x,E_u,E_w,E_aff)
\end{lstlisting}
using affine mappings (\mcode{*}), Minkowski sums (\mcode{+}), and generalized halfspace intersections (\mcode{halfspaceIntersection}) following the procedure from \cite{Bird2023Automatica} as
\vspace{-\baselineskip}
\begin{lstlisting}
nE = size(E_aff,1)
V = [B_u; E_u]*U + [B_w; E_w]*W + [B_aff;zeros(nE,1)]
Y = [A; E_x]*Rk + V
H = eye(nE)
f = E_aff
M = [zeros(nE,Rk.n) eye(nE)]
Rkplus = [eye(Rk.n) zeros(Rk.n,nE)]* halfspaceIntersection(Y,H,f,M)
\end{lstlisting}

\footnotetext{The set \mcode{W} can be generated automatically using the modeling tool Hybrid System DEscription Language (HYSDEL) \cite{Torrisi2004}.}

While MLD and PWA systems are equivalent \cite{Heemels2001}, it may be more computationally-efficient to compute the reachable set of a PWA system directly. Using zonoLAB, state update sets $ \Psi_i $ can be defined as in \eqref{eq:stateUpdateSet} using the linear dynamics over a domain $ D_i $, where $ D_i $ corresponds to the $i^{th}$ convex polytope of the PWA system. For a two-state autonomous system with two regions of linear dynamics, the one-set reachable set can be computed using
\vspace{-\baselineskip}
\begin{lstlisting}
Psi1 = [eye(2);Ad1]*D1 + [zeros(2,1);Bd1]
Psi2 = [eye(2);Ad2]*D2 + [zeros(2,1);Bd2]
Psi = union(Psi1,Psi2)
Rkplus = [zeros(2) eye(2)]*and(Psi,Rk,[eye(2) zeros(2)])
\end{lstlisting}

\begin{figure}
     \centering
     \includegraphics[width=0.85\linewidth]{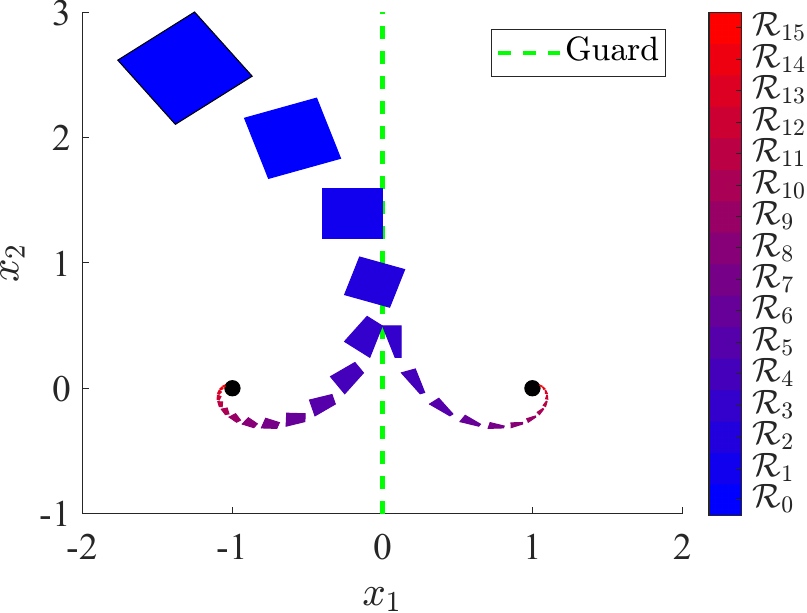}
     \caption{Reachable sets for the two-equilibrium example system from \cite{Bird2023Automatica}, generated by the \mcode{exampleTwoEquilibrium.m} file provided with zonoLAB.
     }
    \label{fig:twoEquExample}
\end{figure}

Fig. \ref{fig:twoEquExample} shows the reachable sets for the two-equilibrium example system from \cite{Bird2023Automatica}. Represented as either an MLD system, generated using HYSDEL \cite{Torrisi2004}, or as a PWA system, the resulting reachable sets computed using the \mcode{exampleTwoEqilibrium.m} file provided with zonoLAB are the same. The reachable sets computed using the PWA representation have slightly more continuous generators and constraints but do not require the definition and bounding of auxiliary variables as in the MLD representation.

\subsection{Closed-loop MPC Analysis}\label{sec-MPC}


Based on the work from \cite{Bird2022ACC}, hybrid zonotopes can exactly represent the piecewise linear feedback control law associated with linear MPC formulations of the form
\begin{equation}\label{eqn-MPC-basic}
\min_{u,x}\|x_N\|^2_{Q_N}+\sum_{k=0}^{N-1}\|x_k\|_{Q}^2+\|u_k\|^2_R, \hspace{30pt} 
\end{equation}
\vspace{-1.5\baselineskip}
\begin{equation*}
\begin{split}
   &\\
     & \:\:\text{s.t. }x_{k+1}=Ax_{k}+Bu_{k}\:,\: u_{k} \in \mathcal{U}\:\forall\:k\in\{0,\dots,N-1\}\:, \\
     & \quad\quad x_{k}\in\mathcal{X}\:\forall\:k\in\{1,\dots,N-1\}\:,\:x_{N}\in\mathcal{X}_{N}\:,
\end{split}
\end{equation*}
where $x_k\in\mathbb{R}^{n_x}$ is the state vector and $u_k\in\mathbb{R}^{n_u}$ is the input vector, related by linear dynamics with $A\in\mathbb{R}^{{n_x}\times {n_x}}$ and $B\in\mathbb{R}^{{n_x}\times {n_u}}$. The sampled state of the system is $x_0$. The inputs, predicted states, and predicted terminal state are constrained to the convex polytopic sets $\mathcal{U}\subset\mathbb{R}^{n_{u}}$, $\mathcal{X}\subseteq\mathbb{R}^{n_{x}}$, and $\mathcal{X}_N\subseteq\mathcal{X}$ and are penalized in the quadratic cost function with weight matrices $R\succ0$, $Q\succeq 0$, and $Q_N\succeq 0$, respectively. To demonstrate this capability, the \mcode{exampleDoubleIntegrator.m} file provided with zonoLAB computes and plots the MPC control law as a hybrid zonotope for the double integrator example from \cite{Bird2022ACC}. 


To perform closed-loop reachability analysis of MPC, the open-loop state-update set $ \Psi $ from \eqref{eq:openLoopStateUpdate} can be combined with the hybrid zonotope representation of the feedback control law (referred to as the \emph{state-input map} in \cite{Siefert2023TAC}). It is assumed that the hybrid zonotope $ \Theta \subset \mathbb{R}^{n+m} $ maps the measured state $ x_k $ to the applied control input $ u_k $ such that $ (x_k, u_k) \in \Theta $ and is defined over a domain $ D(\Theta) $ such that $ \mathcal{R}_k \subseteq D(\Theta) $ for all $ k $. Then, the \emph{closed-loop state-update set} $  \Phi \subset \mathbb{R}^{2n} $ such that $ (x_k,x_{k+1}) \in \Phi $ is computed as
\begin{equation}
    \Phi = \begin{bmatrix}
        \mathbf{I}_{n} & \textbf{0}_{n \times m} & \mathbf{0}_{n} \\
        \mathbf{0}_{n} & \textbf{0}_{n \times m} & \mathbf{I}_{n}
        \end{bmatrix}\big(\Psi \cap_{[\textbf{I}_{n+m}~\textbf{0}_n]} \Theta \big)\:.
\end{equation}
In addition to forward reachability analysis, this closed-loop state-update set can also be used for backward reachability analysis. For example, in \cite{Siefert2022}, backward reachable sets were used to compute the maximal positive invariant set of a closed-loop system under an MPC control law.

To represent MPC formulations beyond the form of \eqref{eqn-MPC-basic}, neural networks can be used to approximate the MPC control policies \cite{Siefert2023TAC,Karg2020}, where representing neural networks as hybrid zonotopes is presented in the following section.



\subsection{Neural Network Representation}\label{sec-NN}


Based on the work from \cite{Ortiz2023}, zonoLAB can exactly represent the input-output mapping of an $L$-layered, ReLU-based, feed-forward, fully-connected neural network $f: \mathbb{R}^n \rightarrow \mathbb{R}^m $ mapping inputs $ x \in \mathbb{R}^n $ to outputs $ y = f(x) \in \mathbb{R}^m $ such that 
\begin{align}
    \begin{split}
        x^0 &= x\:, \\
        x^{\ell+1} &= \phi(W^{\ell} x^{\ell} + b^{\ell})\:, \quad {\ell} \in \{0,\dots,L-1\}\:, \\
        y = f(x) &= W^{L} x^{L} + b^{L}\:,
        \end{split}
\end{align}
where $ W^{\ell} \in \mathbb{R}^{n_{\ell+1} \times n_{\ell}} $ and $ b^{\ell} \in \mathbb{R}^{n_{\ell+1}} $ are the weight matrix and bias vector, respectively, between layers $\ell$ and $\ell+1$, with $n_0 = n$ and $n_{L+1} = m $. All activation functions $\phi$ are ReLU functions that operate element-wise, i.e., for the pre-activation vector $ v^{\ell+1} = W^{\ell} x^{\ell} + b^{\ell} \in \mathbb{R}^{n_{\ell+1}}$, 
\begin{equation}
    \phi(v^{\ell}) = [\varphi(v_1^{\ell}) \; \cdots  \; \varphi(v^{\ell}_{n_{\ell}})]^\top\:,
\end{equation}
where the ReLU function is defined as $\varphi(v_i) = \max\{0,v_i\} $. The key insight from \cite{Ortiz2023} is that a single ReLU activation function $ x_i = \varphi(v_i) $ can be represented as a hybrid zonotope $ \Phi \subset \mathbb{R}^2 $ containing points $ [ v_i \; x_i ]^\top $ over a predetermined domain $ v_i \in [-a, a] $. The user-defined parameter $ a > 0 $ is chosen large enough so that $ |v_i|\leq a $ for all anticipated values of $ v_i $. Specifically, a single ReLU activation function is represented as a hybrid zonotope with $  n_g = 4 $ continuous generators, $ n_b = 1 $ binary generator, and $ n_c = 2 $ constraints. Over the domain $\mathcal{X}$ such that $ x \in \mathcal{X} $, the entire feed-forward ReLU neural network with $ n_N $ ReLU activation functions is exactly represented as a hybrid zonotope $\mathcal{F} $ with $ n_g = n_x + 4 n_N $ continuous generators, $ n_b = n_N $ binary generators, and $ n_c = 3 n_N $ constraints, assuming $\mathcal{X}$ is a zonotope with $ n_x $ generators.

Using zonoLAB, the hybrid zonotope representation of a neural network is computed as 
\vspace{-\baselineskip}
\begin{lstlisting}
[NN,Y] = reluNN(X,Ws,bs,a)
\end{lstlisting}
where \mcode{X} is a \mcode{zono} object with $ n_x $ generators defining the domain $ \mathcal{X}$,  
\mcode{Ws} is a $ 1 \times L $ cell array containing the weight matrices, \mcode{bs} is a $ 1 \times L $ cell array containing the bias vectors, and \mcode{a} defines the domain for each reLU activation function. The output \mcode{NN} is the hybrid zonotope $\mathcal{F} $ such that $ (x,y) \in \mathcal{F} $ and the output \mcode{Y} is the set of outputs from the neural network produced by inputs in the set \mcode{X}.

\begin{figure*}
     \centering
     \includegraphics[width=0.94\linewidth]{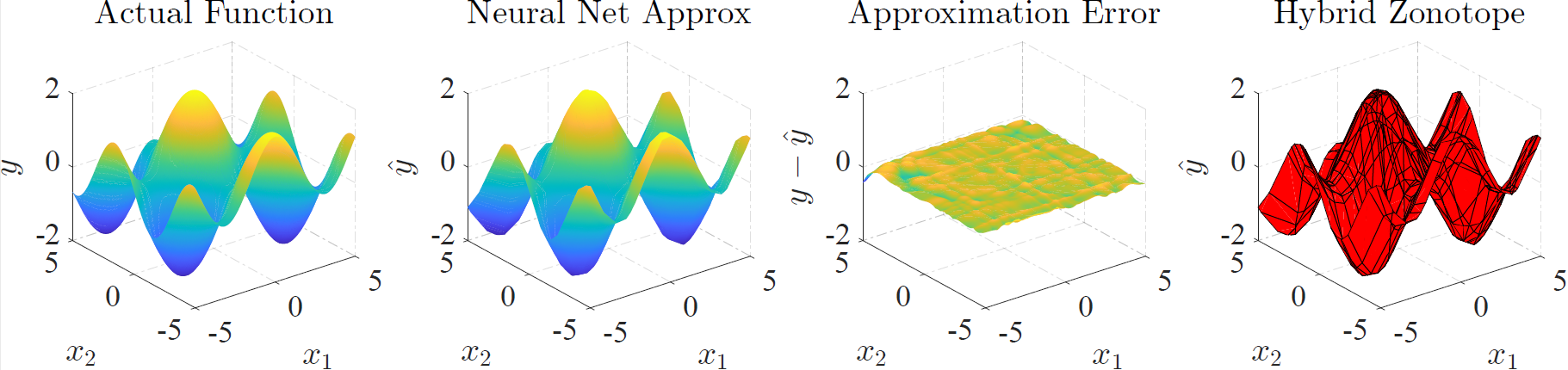}
     \caption{Hybrid zonotope representation of a neural network example from \cite{Ortiz2023} generated by the \mcode{exampleReluNN.m} file provided with zonoLAB.
     }
    \label{fig:NNExample}
\end{figure*}

Fig. \ref{fig:NNExample} shows the nonlinear function $ f(x_1,x_2) = \cos(x_1) + \sin(x_2) $, the approximation of this function using a $ L = 3 $ level neural network with $ n_N = 40 $ total activation functions, the corresponding approximation error, and the exact representation of this neural network as a hybrid zonotope over the domain $ \mathcal{X} = \{ x \in \mathbb{R}^2 \, | \, \|x\|_\infty \leq 5 \} $, which corresponds to one of the examples from \cite{Ortiz2023}.

\begin{figure*}
     \centering
     \begin{subfigure}[t]{0.45\textwidth}
         \includegraphics[width=\linewidth]{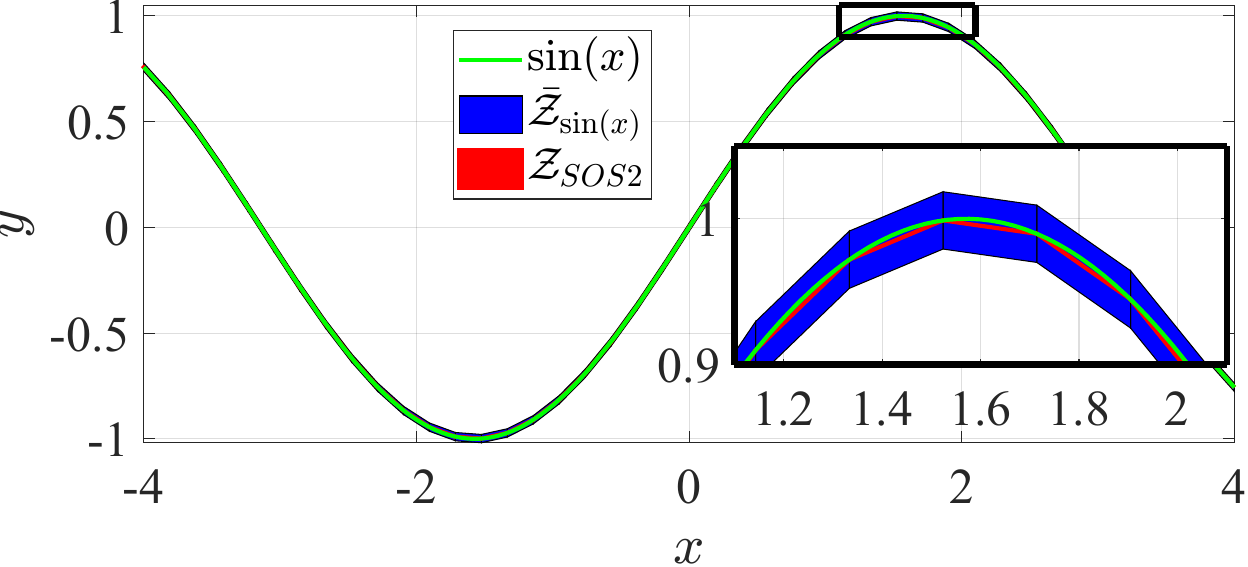}
         \vspace{-15pt}
         \caption{}
     \end{subfigure}
     ~
    \begin{subfigure}[t]{0.22\textwidth}
         \includegraphics[width=\linewidth]{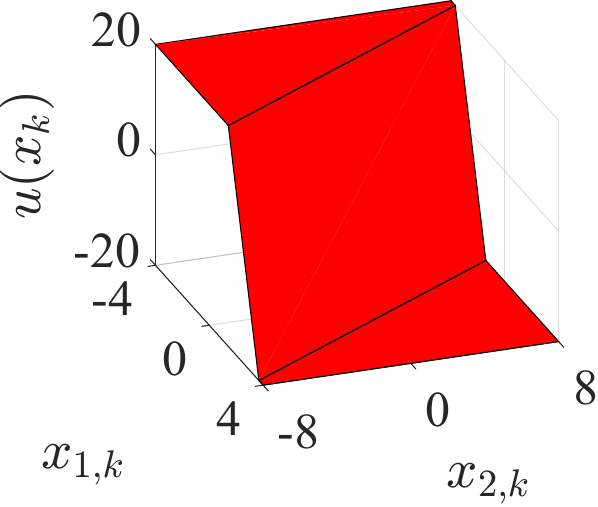}
         \vspace{-15pt}
         \caption{}
     \end{subfigure}
      ~
    \begin{subfigure}[t]{0.29\textwidth}
         \includegraphics[width=\linewidth]{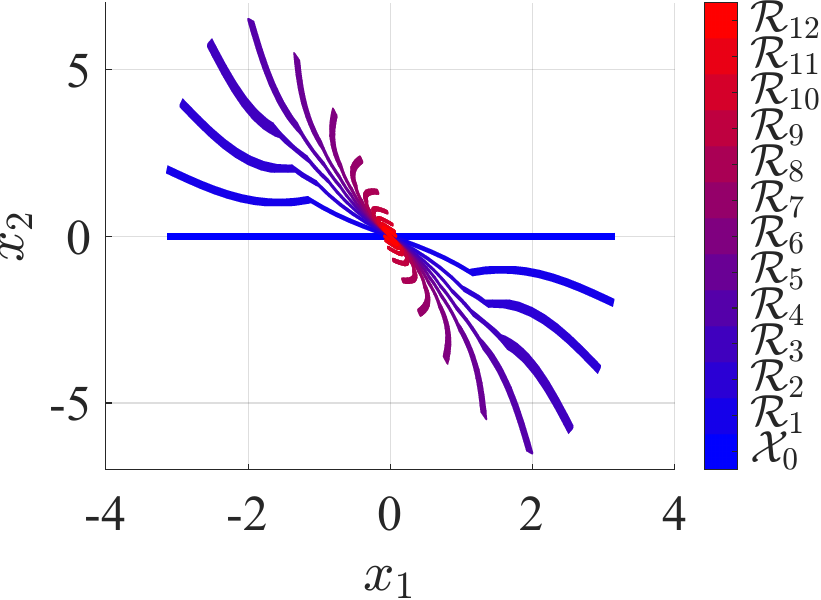}
         \vspace{-15pt}
         \caption{}
     \end{subfigure}
     \caption{Hybrid zonotope representations of a nonlinear function (a), a piecewise linear control law (b), and the resulting closed-loop reachable sets generated by the \mcode{exampleNonlinear.m} file provided with zonoLAB. 
     }
    \label{fig:nonlinExample}
\end{figure*}

\subsection{Nonlinear System Analysis}\label{sec-nonlin}

Based on the work from \cite{Siefert2023ACC,Siefert2023TAC}, zonoLAB can bound nonlinear functions using hybrid zonotopes, which enables reachability analysis of systems with nonlinear dynamics. Considering a unary function $ y = f(x) $ over the domain $ x \in [\underline{x},\overline{x}] \subset \mathbb{R} $, the goal is to find a hybrid zonotope $ \mathcal{Z}_h $ such that $ \{(x,f(x)) \, | \, x \in  [\underline{x},\overline{x}] \} \subseteq \mathcal{Z}_h $. The process used in \cite{Siefert2023ACC,Siefert2023TAC} starts by determining a set of $ n_v $ pairs $ (x_i,f(x_i)), \, i \in \{1,\dots,n_v\} $, such that $ x_1 = \underline{x} $, $ x_{n_v} = \overline{x} $, and $ x_i < x_j, \, \forall \, i < j $. As described in Section \ref{sec:setDefn}, these pairs can be stored in the vertex matrix $ V = [ v_1, \dots, v_{n_v} ] \in \mathbb{R}^{2 \times n_v} $ where $ v_i = [x_i \, f(x_i)]^\top $. The associated incidence matrix $ M \in \mathbb{R}^{n_v \times n_v-1}$ can be defined to create a hybrid zonotope $\mathcal{Z}_v$ representing the union of the $ n_v-1 $ line segments connecting neighboring vertices. While $\mathcal{Z}_v$ approximates the nonlinear function, to determine $ \mathcal{Z}_h $ such that $ \{(x,f(x)) \, | \, x \in  [\underline{x},\overline{x}] \} \subseteq \mathcal{Z}_h $, a set $ \mathcal{E} $ bounding the difference between this piecewise-linear approximation and the nonlinear function must identified to compute $\mathcal{Z}_h = \mathcal{Z}_v \oplus \mathcal{E} $. Processes for automating the choice of $ n_v $ points, computing the error set $ \mathcal{E} $, and using functional decomposition to bound multi-input functions are currently under development \cite{Glunt2024} and will be included in future releases of zonoLAB. While applicable to nonlinear functions in general, this approach can also be used to determine hybrid zonotopes that bound 1) the open-loop state update set $ \Psi $ from Section \ref{sec:setBasedMap} for a system with nonlinear dynamics $ x_{k+1} = f(x_k, u_k) $, 2) nonlinear feedback control laws, and 3) input-output mappings for neural networks with nonlinear activation functions.

\begin{table}
\caption{Hybrid zonotope set representation complexity, number of convex sets, and plotting time for each reachable set shown in Fig. \ref{fig:nonlinExample}(c).}
\centering
\begin{tabular}{lrrrrr}
   \hline
   & $n_g$ & $n_b$ & $n_c$ & $|\mathcal{T}|$ & 
   $t_{plot} $ (s) \\ \hline
$\mathcal{R}_0$ & 2 & 0 & 0 & 0 & 0.2   \\
$\mathcal{R}_1$ & 103 & 44 & 59 & 38 & 1.1   \\
$\mathcal{R}_2$ & 204 & 88 & 118 & 108 & 3.6   \\
$\mathcal{R}_3$ & 305 & 132 & 177 & 162 & 6.8   \\
$\mathcal{R}_4$ & 406 & 176 & 236 & 226 & 12.4   \\
$\mathcal{R}_5$ & 507 & 220 & 295 & 300 & 20.1   \\
$\mathcal{R}_6$ & 608 & 264 & 354 & 384 & 29.2   \\
$\mathcal{R}_7$ & 709 & 308 & 413 & 518 & 44.9   \\
$\mathcal{R}_8$ & 810 & 352 & 472 & 672 & 62.9   \\
$\mathcal{R}_9$ & 911 & 396 & 531 & 984 & 95.0   \\
$\mathcal{R}_{10}$ & 1012 & 440 & 590 & 1568 & 186.2  \\ 
$\mathcal{R}_{11}$ & 1113 & 484 & 649 & 2666 & 292.3  \\ 
$\mathcal{R}_{12}$ & 1214 & 528 & 708 & 4732 & 482.3  \\ 
\end{tabular}
\label{tab:nonlinExample}
\end{table}

To demonstrate the ability to bound nonlinear functions and conduct closed-loop reachability analysis of nonlinear dynamic systems, Fig. \ref{fig:nonlinExample} shows numerical results based on the example from \cite{Siefert2023ACC}, which is provided with zonoLAB in the \mcode{exampleNonlinear.m} file. This examples considers a two-state, one-input, discrete-time nonlinear system (where the dynamics include $ \sin(x) $) under feedback control based on a saturated Linear Quadratic Regulator (LQR). Fig. \ref{fig:nonlinExample}(a) shows the ability to approximate the $ \sin(x) $ function using $\mathcal{Z}_v $ and bound the function using $ \mathcal{Z}_h $, which is then used to produce the open-loop state update set $ \Psi $. Fig. \ref{fig:nonlinExample}(b) shows the saturated LQR control law represented as a hybrid zonotope, which is combined with $ \Psi $ to create the closed-loop state update set $ \Phi $. Fig. \ref{fig:nonlinExample}(c) shows the forward reachable sets of the closed-loop system for twelve time steps. Finally, for each reachable set, Table \ref{tab:nonlinExample} shows $ n_g $, $ n_b $, $ n_c $, $ |\mathcal{T}| $, and the computation time (on a standard desktop computer) to determine $\mathcal{T} $ and plot the resulting constrained zonotopes, using the process described in Section \ref{sec:plotting}. Despite the large number of binary factors, it is relatively quick to identify the number of non-empty sets using GUROBI and to plot each set using linear programming to find each vertex and face. However, these results motivate the need for set representation complexity reduction based on methods from \cite{Bird2022Dissertation}, which will be provided in future releases of zonoLAB.



\section{Conclusions}\label{sec-conclusions}

This paper introduced zonoLAB as a MATLAB toolbox using zonotopic set representations, including zonotopes, constrained zonotopes, and hybrid zonotopes, for the analysis of dynamic systems and their controllers. The zonoLAB toolbox provides a suite of computationally-efficient set-based operations that exploit the specific structures of these zonotopic set representations and user-friendly methods that package complex analysis into simple function calls. The use of hybrid zonotopes in particular provides a scalable approach to applications where sets of interest correspond to the union of convex polytopes. The zonoLAB toolbox currently includes the basic functionality for defining zonotopic sets and computing fundamental set operations along with several illustrative examples. Ongoing efforts aim to expand the capability of zonoLAB in areas such as representing neural networks with a variety of activation functions, functional decomposition methods for bounding nonlinear functions with hybrid zonotopes, and complexity reduction techniques for zonotopic sets.





\bibliographystyle{IEEEtran}
\bibliography{00_Main}

\end{document}